\documentclass[preprint,12pt]{elsarticle}

\usepackage{amssymb}

\usepackage{fullpage}
\usepackage{float}
\restylefloat{table}

\usepackage{amsmath}
\usepackage{graphicx}
\usepackage[dvipsnames]{xcolor}
\usepackage[colorlinks]{hyperref}
\def\bR{{\mathbf{R}}}
\def\bx{{\mathbf{x}}}

\usepackage{comment}
\usepackage{epstopdf}

\journal{arXiv}

\begin{document}

\begin{frontmatter}



\title{M-SPARC: \textsc{Matlab}-Simulation Package for Ab-initio Real-space Calculations}


\author{Qimen Xu, Abhiraj Sharma, and Phanish Suryanarayana}

\address{College of Engineering, Georgia Institute of Technology, Atlanta, GA 30332, USA}

\begin{abstract}
We present M-SPARC: \textsc{Matlab}-Simulation Package for Ab-initio Real-space Calculations. It can perform pseudopotential spin-polarized and unpolarized Kohn-Sham Density Functional Theory (DFT) simulations for  isolated systems such as molecules as well as extended systems such as crystals, surfaces, and nanowires. M-SPARC provides a rapid prototyping platform for the development and testing of new algorithms and methods in real-space DFT, with the potential to significantly accelerate the rate of advancements in the field. It also provides a convenient avenue for the accurate first principles study of small to moderate sized systems.
\end{abstract}


\end{frontmatter}

\section*{Code Metadata}
\label{sec:metadata}

\begin{table*}[h!]
\centering

\begin{tabular}{|l|p{6.5cm}|p{6.5cm}|}
\hline
C1 & Current code version & v1.0.0 \\
\hline
C2 & Permanent link to code/repository used for this code version & $https://github.com/SPARC-X/M-SPARC$\\
\hline
C3 & Code Ocean compute capsule & \\
\hline
C4 & Legal Code License   & GNU General Public License v3.0 \\
\hline
C5 & Code versioning system used & git \\
\hline
C6 & Software code languages, tools, and services used & \textsc{Matlab 2013+}\\
\hline
C7 & Compilation requirements, operating environments \& dependencies & OS: Unix, Linux, MacOS, or Windows \\
\hline
C8 & If available Link to developer documentation/manual & $https://github.com/SPARC-X/M-SPARC/tree/master/doc$ \\
\hline
C9 & Support email for questions & phanish.s@gmail.com \\
\hline
\end{tabular}
\caption{Code metadata}
\label{metadata} 
\end{table*}








\section{Motivation and significance}
\label{sec:motivation}
Kohn-Sham Density Functional Theory (DFT) \cite{hohenberg1964inhomogeneous,Kohn1965} is one of the most widely used electronic structure theories for understanding and predicting materials properties from the first principles of quantum mechanics \cite{becke2014perspective,pribram2015dft}, without the need for any empirical or adjustable parameters. The popularity of DFT  can be attributed to its high accuracy to cost ratio when compared to other such ab initio methods. However the efficient solution of the Kohn-Sham problem remains a significant challenge, which limits the size and length scales that are accessible to such a rigorous first principles investigation. 

The computational cost and memory associated with diagonalization based solutions of the Kohn-Sham equations scale cubically and quadratically with respect to system size, respectively \cite{Martin2004}. The accompanying prefactors are particularly large when systematically improvable discretizations are used. Though the Kohn-Sham problem can be reformulated in terms of the truncated density matrix to develop linear scaling methods \cite{Goedecker,Bowler2012}, they suffer from a number of limitations. These  include being restrictive in the types of systems that can be studied and having a significantly larger prefactor, which makes them inefficient relative to their cubic scaling counterparts for small to moderate system sizes \cite{suryanarayana2018sqdft,suryanarayana2017nearsightedness}.  

In view of the above bottlenecks, nearly all established DFT codes take advantage of the efficiency and parallel scalability provided by lower level programming languages such as C, C++, and Fortran. In particular, it is common to employ multiple levels of parallelization, the nature of which also changes between different parts of the code. This leads to significant code complexity, making the testing of new ideas and methods prone to error, while requiring excessively large effort for implementation. As a consequence, the rate of advancements in the field can be significantly  hindered. This motivates the development of simple but accurate codes written in high level programming languages such as Python and \textsc{Matlab}, which enable rapid prototyping.  

\textsc{KSSOLV} \cite{yang2009kssolv} is one such \textsc{Matlab} code for the planewave method, traditionally the discretization of choice in Kohn-Sham DFT \cite{VASP,giannozzi2009quantum,CASTEP,ABINIT,gygi2008architecture,valiev2010nwchem}. However, to the best of our knowledge, no such counterpart exists for real-space methods, which have gained significant attention recently \cite{bernholc1991structural,chelikowsky1994finite,seitsonen1995real,gygi1995real,briggs1996real,iwata2010massively,ghosh2017sparc1,ghosh2017sparc2}, in part due to their high scalability for large-scale parallel computing \cite{hasegawa2011first,ghosh2017sparc1,ghosh2017sparc2}, flexibility in the choice of boundary conditions \cite{natan2008real,Phanish2012,ghosh2019symmetry}, and amenability to the development of linear scaling methods \cite{osei2014accurate,suryanarayana2018sqdft}. Motivated by this, in this work we develop M-SPARC: \textsc{Matlab}-Simulation Package for Ab-initio Real-space Calculations. It provides a rapid prototyping platform for the development and testing of new algorithms and methods in real-space DFT. Additionally, it provides a convenient avenue for the accurate first principles study of small to moderate sized systems.


\section{Software description} 
\label{sec:description}

The central focus of M-SPARC is the solution of the Kohn-Sham equations \cite{Kohn1965}:
\begin{equation} \label{Eqn:KSeq}
\left( \mathcal{H}^{\sigma} \equiv - \, \frac{1}{2} \nabla^2 + V_{\rm eff}^{\sigma} \left[ \rho^{\alpha}, \rho^{\beta}; \bR \right] \right)  \psi_n^{\sigma}  =  \lambda_n^{\sigma} \psi_n^{\sigma} \,, \quad n=1,2, \ldots, N_s \,, \quad \sigma \in \{\alpha, \beta\} \,,
\end{equation}
where $\mathcal{H}^{\sigma}$ is the Hamiltonian, $\psi_n^\sigma$ are the Kohn-Sham orbitals with energies $\lambda_n^{\sigma}$, $V_{\rm eff}^{\sigma}$ is the effective potential, $N_s$ is the number of states, $\sigma$ denotes the spin component, and $\bR$ represents the collection of atomic positions. In addition, $\rho^{\alpha}$ and $\rho^{\beta}$ are the spin-up and spin-down components of the electron density, respectively: 
\begin{equation} \label{Eqn:SpinElectronDensity}
\rho^{\sigma}(\bx) = \sum_{n=1}^{N_s}  g_n^{\sigma} {|\psi_n^{\sigma}(\bx)|}^2 \,, \quad \sigma \in \{\alpha,\beta\} \,, \quad \bx \in \mathbb{R}^3
\end{equation}
where $g_n^{\sigma}$ are the orbital occupations, typically given by the Fermi-Dirac distribution. In computations, zero Dirichlet and Bloch-periodic boundary conditions are prescribed on the orbitals along the directions in which the system is  finite and extended, respectively. Note that when spin is neglected, i.e., for spin-unpolarized calculations, all quantities become independent of the spin, whereby the two coupled nonlinear eigenproblems in Eqn.~\ref{Eqn:KSeq} reduce to a single nonlinear eigenproblem.  


\subsection{Software Architecture }
\label{subsection:architecture}
\begin{figure}[h!]
\centering
\includegraphics[keepaspectratio=true,width=0.65\textwidth]{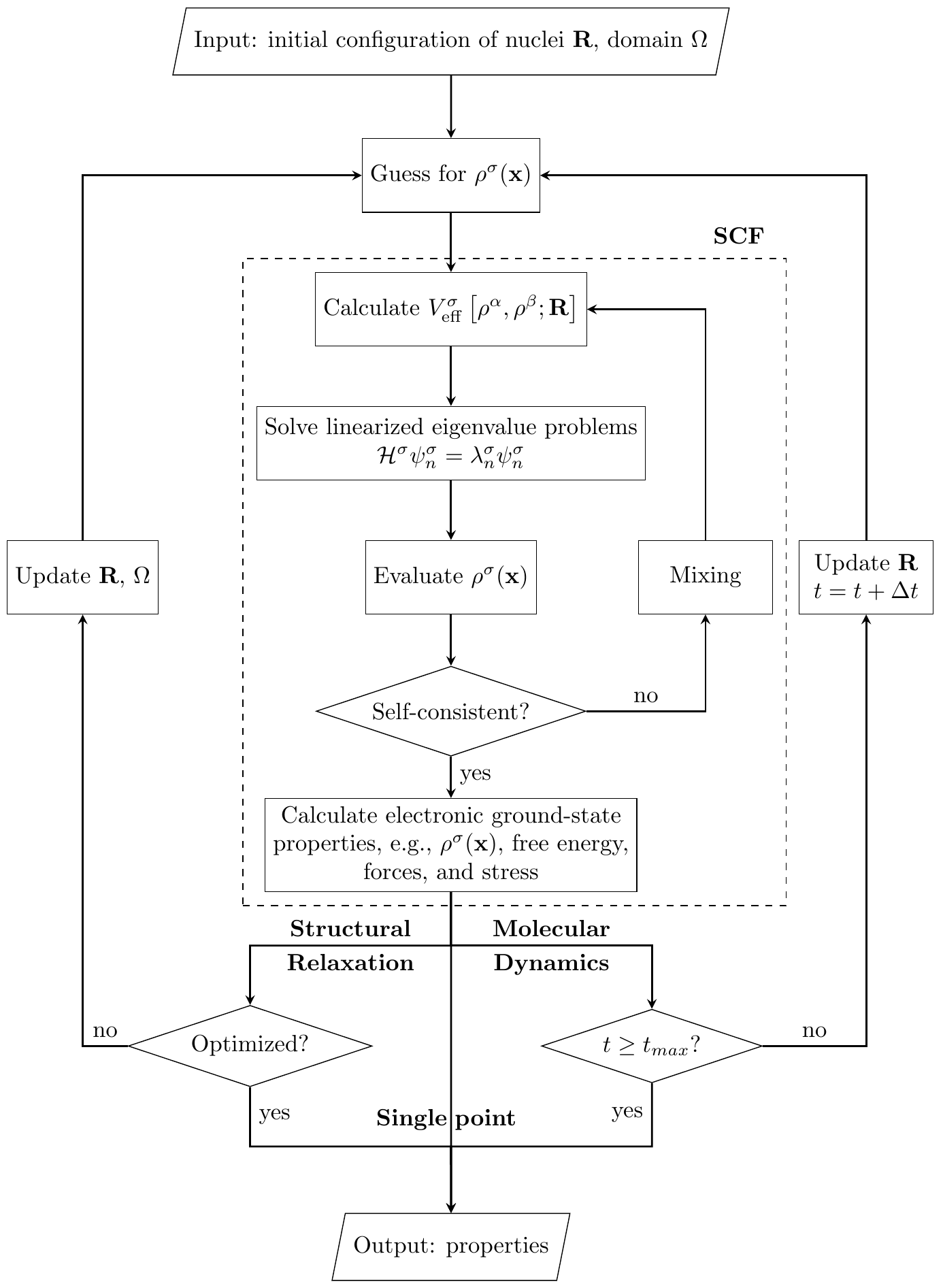}
\caption{Overview of the M-SPARC framework for performing DFT calculations.}
\label{Fig:OverallFlowChart}
\end{figure}

M-SPARC is written exclusively using the \textsc{Matlab} language. It employs a high-order central finite-difference approximation for discretization of the equations and a trapezoidal rule for spatial integrations.  A pictorial overview of the M-SPARC framework for performing DFT calculations is illustrated in Fig.~\ref{Fig:OverallFlowChart}. It requires two input files: (i) \texttt{.inpt} file containing user options and parameters, including the dimensions of the cell, boundary conditions, information about the finite-difference grid, and choice of exchange-correlation functional; and (ii) \texttt{.ion} file containing the atomic information, including the atom type, its spatial position, and path to its pseudopotential file.

Three types of calculations can be performed in M-SPARC: single point, structural relaxation, and molecular dynamics. In single point calculations, the electronic ground-state is calculated for a given atomic configuration. In structural relaxation simulations, the energy is minimized with respect to the atomic positions or volume of the cell, while employing the computed Hellmann-Feynman atomic forces \cite{ghosh2017sparc1,ghosh2017sparc2} or stress tensor \cite{sharma2018calculation}, respectively. In molecular dynamics simulations, the ionic positions, velocities, and accelerations are evolved by integrating the equations of motion, while utilizing the atomic forces. Indeed, the electronic ground state needs to be determined for every atomic configuration encountered during the structural relaxation and molecular dynamics simulations.

The electronic ground state is determined using the self-consistent field (SCF) method \cite{Martin2004}. Specifically, a fixed-point iteration is performed with respect to either the electron density or the potential. For the very first electronic ground state calculation, the superposition of isolated atom electron densities is used as initial guess for the electron density, whereas for every subsequent such calculation, extrapolation based on previous solutions is used \cite{alfe1999ab}. The convergence of the SCF iteration is accelerated using the restarted variant of the Periodic Pulay mixing scheme \cite{pratapa2015restarted,banerjee2016periodic}, with the option of real-space preconditioning \cite{kumar2019preconditioning}. For spin-polarized calculations, mixing is performed simultaneously for both spin components, i.e., using a vector of twice the original length containing both spin-up and spin-down density/potential components. The SCF iteration is considered to be converged when self-consistency in the solution is achieved.

In each SCF iteration, partial diagonalization of the linearized eigenproblem is performed using the CheFSI method \cite{zhou2006self,zhou2006parallel}, with multiple Chebyshev filtering steps in the very first iteration of the entire DFT simulation \cite{zhou2014chebyshev}.  Zero Dirichlet and Bloch-periodic boundary conditions are prescribed on the orbitals $\psi_n^{\sigma}$ along the directions in which the system is finite and extended, respectively. While performing the Hamiltonian-matrix products, the Kronecker product formulation for the Laplacian \cite{sharma2018real} is used, with the remaining terms handled in a matrix-free fashion. A local formulation of the electrostatics in terms of ionic pseudocharges is employed \cite{Suryanarayana2014524,ghosh2014higher}, wherein the electrostatic potential (component of $V_{\rm eff}^{\sigma}$) is determined by the solution of the Poisson equation. Dirichlet and periodic boundary conditions are prescribed on the electrostatic potential along directions in which the system is finite and extended, respectively. The Dirichlet boundary condition values are determined using a multipole expansion for isolated systems and a dipole correction for surfaces and nanowires \cite{burdick2003parallel,natan2008real}. The linear system is solved using the AAR method \cite{pratapa2016anderson,suryanarayana2019alternating} in conjunction with Cholesky preconditioning.  

After completion of the DFT simulation, in addition to the parameters used in the calculation, all quantities of interest such as the orbitals, occupations, electron density, and electrostatic potential are stored in the structure denoted by \texttt{S}. General information such as input parameters, progress of the SCF iteration, energy, maximum force, and timing are written into the \texttt{.out} file. Additionally, \texttt{.static}, \texttt{.geopt}, \texttt{.cellopt}, and \texttt{.aimd} files are written for single point, atomic structural relaxation, cell structural relaxation, and molecular dynamics simulations, respectively. The \texttt{.static} file contains the atomic positions and forces; the \texttt{.geopt} file contains the atomic positions and forces for each atomic relaxation step; the \texttt{.cellopt} contains the cell information and stress tensor for each cell relaxation step; and the \texttt{.aimd} file contains atomic positions, forces, and velocities. Note that a \texttt{.restart} file is also written for structural relaxations and molecular dynamics, which can be used to restart the simulation. 

\subsection{Software Functionalities}
\label{subsection:functionalities}
M-SPARC can perform spin-polarized and unpolarized  ab initio calculations based on pseudopotential Kohn-Sham DFT for isolated systems such as molecules as well as extended systems such as crystals, surfaces, and nanowires. Specifically, it can currently perform single point calculations for a given atomic configuration, structural relaxations with respect to atomic positions or cell volume, and NVE molecular dynamics simulations, all using either the ONCV \cite{hamann2013optimized}  or Troullier-Martin pseudopotentials \cite{Troullier} and either the LDA \cite{PhysRevB.23.5048,perdew1986accurate} or GGA \cite{perdew1996generalized} exchange-correlation functionals. In so doing, M-SPARC can calculate the energy of the system as well as the Hellmann-Feynman atomic forces and stress tensor. The output from such DFT calculations can be used to calculate a number of properties, such as equilibrium bond lengths, HOMO-LUMO gap, dipole moment, surface energy, cohesive energy, defect energy, lattice constant, elastic moduli, density of states, electronic band structure, and pair correlation function. 


\section{Illustrative Examples}
\label{sec:examples}

\begin{figure}[h!]
\centering
\includegraphics[keepaspectratio=true,width=0.98\textwidth]{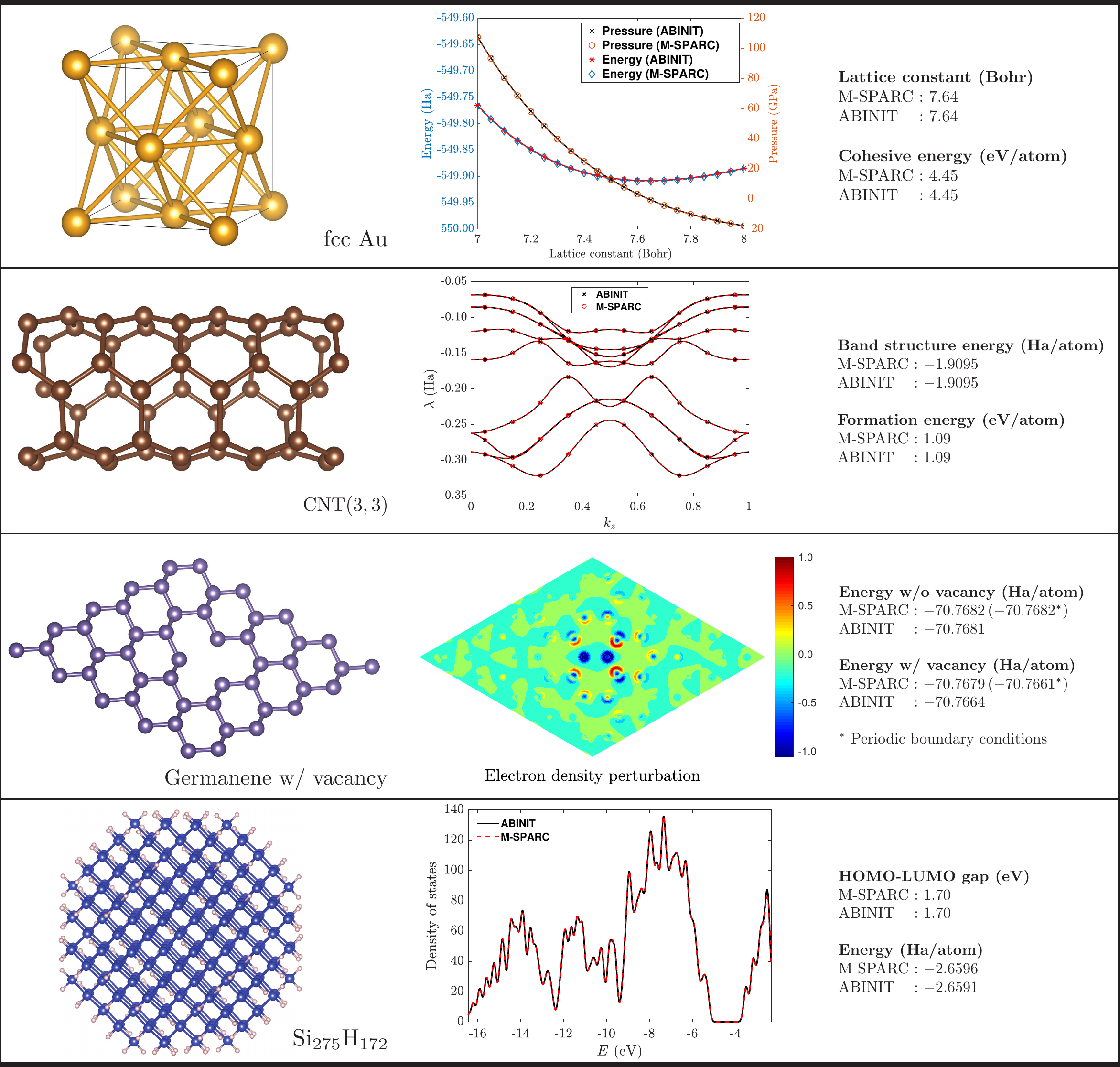}
\caption{Examples demonstrating some of the major functionalities of M-SPARC.}
\label{Fig:Examples}
\end{figure}

We now demonstrate some of the major functionalities of M-SPARC through representative examples. Specifically, we consider a (i) 4-atom cell of bulk fcc gold with LDA, $7 \times 7 \times 7$ grid for Brillouin zone integration, and mesh-size of 0.3 Bohr; (ii) 49-atom cell of germanene ($5 \times 5$ supercell with a vacancy) with GGA, $2 \times 2 $ grid for Brillouin zone integration, and mesh-size of 0.3 Bohr; (iii) 12-atom (3,3) carbon nanotube with GGA, $10$ points for Brillouin zone integration, and mesh-size of 0.3 Bohr; and (iv) a $\mathrm{Si}_{275}\mathrm{H}_{172}$ nanodot with LDA and mesh-size of 0.5 Bohr. We employ ONCV pseudopotentials \cite{hamann2013optimized} with 19, 14, 4, 4, and 1 valence electrons for gold, germanium, carbon, silicon, and hydrogen, respectively. In all cases, we use the $12$-th order finite-difference approximation for discretization of the equations. Input files for these examples are available as part of the distribution. We present the results so obtained in Fig.~\ref{Fig:Examples} and compare them to highly converged results obtained by the established planewave code ABINIT \cite{ABINIT}. It is clear that that there is very good agreement between M-SPARC and ABINIT, verifying the accuracy of the implementation. Indeed, the agreement further increases as the discretization is refined in M-SPARC, i.e., the mesh-size is made smaller.  \vspace{5mm}
 
\section{Impact}
\label{sec:impact}
%
%
%
%
%

M-SPARC provides a rapid prototyping platform for the development and testing of new algorithms/methods in Kohn-Sham DFT, particularly for  the real-space method. This is evidenced by the number of accurate and efficient methods developed in M-SPARC that have resulted in immediate publications, including the linear scaling Spectral Quadrature (SQ)  DFT method that is identically applicable to insulating and metallic systems \cite{pratapa2015spectral}; coarse-grained DFT formulation that enables the study of crystal defects at realistic concentrations \citep{Phanish2012}; cyclic and helical symmetry-adapted DFT formulations that allow for the highly efficient study of systems with such symmetries, enabling the ab initio study of nanomaterials subjected to bending and torsional deformations \cite{banerjee2016cyclic,banerjee2018ab}; Kronecker product formulation of the kinetic energy operator that reduces the cost of real-space DFT for non-orthogonal systems; the Discontinuous Discrete Basis Projection (DDBP) method that significantly increases the efficiency of real-space DFT by reducing the size of the Hamiltonian by up to three orders of magnitude  \cite{xu2018discrete}; real-space formulation for the Hellmann-Feynman stress tensor in DFT \cite{sharma2018calculation}; real-space formulation for isotropic Fourier-space preconditioners that can accelerate the SCF iteration in DFT \cite{kumar2019preconditioning}.

M-SPARC and its variants are currently being used by multiple research groups. Moving forward, the user base is expected to significantly expand, given the current open source software release of M-SPARC and the noticeable emphasis placed on the development of real-space DFT \cite{bernholc1991structural,chelikowsky1994finite,seitsonen1995real,gygi1995real,briggs1996real,iwata2010massively,ghosh2017sparc1,ghosh2017sparc2}. Possible avenues for M-SPARC to have an immediate impact in real-space DFT include:  implementation of sophisticated exchange-correlation functionals such as hybrids;  preconditioners for accelerating the convergence of eigensolvers; mixing schemes and associated preconditioners for accelerating SCF convergence, particularly for spin polarized calculations; techniques for reducing the size of the Hamiltonian by projection onto a significantly smaller basis; novel boundary conditions that accurately and efficiently capture the physics/chemistry of the system; formulations for reducing the eggbox effect; and machine learning models in the context of DFT. Note that even though the main utility of M-SPARC is in rapid prototyping for large-scale implementations, it provides a convenient avenue for the ab initio investigation of small to moderately sized systems, wherein the time to solution (i.e., wall time) is only a few fold larger than established parallel DFT codes.  


\section{Conclusions}
\label{}
M-SPARC is now a mature \textsc{Matlab} code for performing real-space Kohn-Sham DFT calculations, motivating its open-source release along with this article. It provides a rapid prototyping platform for the development and testing of new algorithms and methods in Kohn-Sham DFT, particularly for the real-space method. As a result, M-SPARC has the potential to significantly accelerate the rate of advancements in the field, which can enable a number of new and exciting applications in science and engineering that were previously intractable.


\section{Conflict of Interest}
We wish to confirm that there are no known conflicts of interest associated with this publication and there has been no significant financial support for this work that could have influenced its outcome.


\section*{Acknowledgements}
The work was supported by the grant DE-SC0019410 funded by the U.S. Department of Energy, Office of Science. Initial development efforts were supported by the grants 1333500 and 1553212 funded by the U.S. National Science Foundation. 



\bibliographystyle{elsarticle-num} 

\clearpage

\end{document}